\documentclass[10pt,journal,final,finalsubmission,twocolumn]{IEEEtran}

\usepackage{graphicx}

\usepackage{amsmath}   
\usepackage{amsfonts}

\newtheorem{theorem}{\textbf{Theorem}}
\newtheorem{lemma}{\textbf{Lemma}}

\begin{document}
%
\title{How Does Multiple-Packet Reception Capability Scale the Performance of Wireless Local Area Networks?}
%
%
\author{Ying~Jun~(Angela)~Zhang,~\IEEEmembership{Member,~IEEE,}
        Peng~Xuan~Zheng,~\IEEEmembership{Student Member,~IEEE,}
        and~Soung~Chang~Liew,~\IEEEmembership{Senior Member,~IEEE}
\thanks{This work is supported in part by the Competitive Earmarked Research Grant (Project number 418506, 418707, and 414106) established under the University Grant Committee of Hong Kong.}
\thanks{Ying Jun Zhang and Soung Chang Liew are with Department of Information Engineering, The Chinese University of Hong Kong, Hong Kong. Email:\{yjzhang, scliew\}@ie.cuhk.edu.hk.}
\thanks{Peng Xuan Zheng is now with Department of Computer Science, Purdue University. Email: pzheng@cs.purdue.edu. He was with Department of Information Engineering, The Chinese University of Hong Kong, Hong Kong.}
}
%
%
%



\maketitle

\begin{abstract}
 Thanks to its simplicity and cost efficiency, wireless local area network (WLAN) enjoys unique advantages in providing high-speed and low-cost wireless services in hot spots and indoor environments. Traditional WLAN medium-access-control (MAC) protocols assume that only one station can transmit at a time: simultaneous transmissions of more than one station cause the destruction of all packets involved. By exploiting recent advances in PHY-layer multiuser detection (MUD) techniques, it is possible for a receiver to receive multiple packets simultaneously. This paper argues that such multipacket reception (MPR) capability can greatly enhance the capacity of future WLANs. In addition, the paper provides the MAC-layer and PHY-layer designs needed to achieve the improved capacity. First, to demonstrate MPR as a powerful capacity-enhancement technique, we prove a ``super-linearity" result, which states that the system throughput per unit cost increases as the MPR capability increases. Second, we show that the commonly deployed binary exponential backoff (BEB) algorithm in today's WLAN MAC may not be optimal in an MPR system, and that the optimal backoff factor increases with the MPR capability, the number of packets that can be received simultaneously. Third, based on the above insights, we design a joint MAC-PHY layer protocol for an IEEE 802.11-like WLAN that incorporates advanced PHY-layer signal processing techniques to implement MPR.
\end{abstract}


\section{Introduction}
%
%
%
%
\subsection{Motivation}
The last decade has witnessed a surge of interest in wireless local area networks (WLAN), where mobile stations share a common wireless medium through contention-based medium access control (MAC). In WLANs, collision of packets occurs when more than one station transmits at the same time, causing a waste of bandwidth. Recent advances in multiuser detection (MUD) techniques \cite{Verdu:98} open up new opportunities for resolving collisions in the physical (PHY) layer. For example, in CDMA \cite{Rapajic:99} or multiple-antenna \cite{Telatar:99} systems, multiple packets can be received simultaneously using MUD techniques without collisions. It is expected that, with improved multipacket reception (MPR) capability from the PHY layer, the MAC layer will behave differently from what is commonly believed. In particular, to fully utilize the MPR capability for capacity enhancement in WLAN, it is essential to understand the fundamental impact of MPR on the MAC-layer design. As such, this paper is an attempt to study the MAC-layer throughput performance and the collision resolution schemes for WLANs with MPR.

\subsection{Key Contributions}
The key contributions of this paper are as follows:
\begin{itemize}
  \item {To demonstrate MPR as a powerful capacity-enhancement technique at the system level, we analyze the MAC-layer throughput of WLANs with MPR capability under both finite-node and infinite-node assumptions. Our model is sufficiently general to cover both carrier-sensing and non-carrier-sensing networks. We prove that in random-access WLANs, network throughput increases super-linearly with the MPR capability of the channel. That is, throughput divided by \emph{M} increases as \emph{M} increases, where \emph{M} is the number of packets that can be resolved simultaneously. The super-linear throughput scaling implies that the achievable throughput per unit cost increases with MPR capability of the channel. This provides a strong incentive to deploy MPR in next-generation wireless networks.}
  \item {We study the effect of MPR on the MAC-layer collision resolution scheme, namely exponential backoff (EB). When packets collide in WLANs, an EB scheme is used to schedule the retransmissions, in which the waiting time of the next retransmission will get multiplicatively longer for each collision incurred. In the commonly adopted binary exponential backoff (BEB) scheme (e.g., used in Ethernet \cite{Ansi}, WiFi \cite{WLAN}, etc.), the multiplicative (a backoff factor) is equal to 2. We show in this paper that the widely used BEB does not necessarily yield the close-to-optimal network throughput with the improved MPR capability from the PHY layer. As a matter of fact, BEB is far from optimum for both non-carrier-sensing networks and carrier-sensing networks operated in basic access mode. The optimal backoff factor increases with the MPR capability. Meanwhile, BEB is close to optimum for carrier-sensing networks when RTS/CTS access mode is adopted.}
  \item {Built on the theoretical underpinnings established above, we propose a practical protocol to fully exploit the MPR capability in IEEE 802.11-like WLANs. In contrast to \cite{Zhao:03}-\cite{Zhao:04}, we consider not only the MAC layer protocol design, but also the PHY-layer signal processing to enable MPR in distributed random-access WLANs. As a result, the proposed protocol can be implemented in a fully distributed manner with marginal modification of current IEEE 802.11 MAC.}
\end{itemize}

\subsection{Related Work on MPR and Collision Resolution Schemes}
The first attempt to model a general MPR channel in random-access wireless networks was made by Ghez, Verdu, and Schwartz in \cite{Ghez:88}-\cite{Ghez:89} in 1988 an 1989, respectively, in which stability properties of conventional slotted ALOHA with MPR were studied under a simple infinite-user and single-buffer assumption. No collision resolution scheme (such as EB) was considered therein. This work was extended to CSMA systems by Chan et al in \cite{Chan:04} and to finite user ALOHA systems by Naware et al in \cite{Naware:05}. It has been shown in \cite{Ghez:88}-\cite{Naware:05} that MPR improves the stable throughput of ALOHA only when the MPR capability is comparable to the number of users in the system. In practical networks where the MPR capability is much smaller than the number of users, the stable throughput of conventional ALOHA is equal to 0, same as the case without MPR. To date, little work has been done to investigate the throughput enhancing capability of MPR in practical WLANs with collision resolution schemes. Our paper here is an attempt along this direction.

Protocols that exploit the MPR capability of networks have been studied by Zhao and Tong in \cite{Zhao:03}-\cite{Zhao:04}. In \cite{Zhao:03}, a multi-queue service room (MQSR) MAC protocol was proposed for networks with heterogeneous users. The drawback of the MQSR protocol is its high computational cost due to updates of the joint distribution of all users' states. To reduce complexity, a suboptimal dynamic queue protocol was proposed in \cite{Zhao:04}. In both protocols, access to the common wireless channel is controlled by a central controller, which grants access to the channel to an appropriate subset of users at the beginning of each slot. In \cite{Chan:05}, Chan et al proposed to add a MUD layer to facilitate MPR in IEEE 802.11 WLAN. To implement the MUD techniques mentioned as examples in \cite{Chan:05}, the AP is assumed to have perfect knowledge of the number of concurrent transmissions, the identities of the transmitting stations, and the channel coefficients. These information, while easy to get in a network with centralized scheduling (e.g., cellular systems), is unkown to the AP a priori in random access networks. Moreover, the preambles of concurrent packets overlap, and hence it is difficult for the AP to have a good estimation of the channel coefficients with the current protocol. By contrast, our paper provides a solution to this issue by incorporating blind signal processing in the proposed  protocol.

Exponential Backoff (EB) as a collision resolution technique has been extensively studied in different contexts \cite{Goodman:88}-\cite{Kwak:05}. Stability upper bound of BEB has been given by Goodman under a finite-node model in \cite{Goodman:88} and recently improved by Al-Ammal in \cite{Ammal:01}. The throughput and delay characteristics of a slightly modified EB scheme have been studied in \cite{Jeong:95} in the context of slotted ALOHA. The characteristics of EB in steady state is further investigated in \cite{Kwak:05} in time slotted wireless networks with equal slot length. All the existing work on EB has assumed that the wireless channel can only accommodate one ongoing transmission at a time. This paper is a first attempt to look at EB for an MPR system.

The remainder of this paper is organized as follows. In Section II, we describe the system model and introduce the background knowledge on MUD and EB. In Section III, we prove that the maximum achievable throughput of MPR WLAN scales super-linearly with the MPR capability of the channel. In Section IV, the effect of MPR on EB is investigated. We show that the widely used BEB scheme is no longer close-to-optimal in MPR networks. To realize MPR in IEEE 802.11 WLANs, a MAC-PHY protocol is presented in Section V. In Section VI, we discuss some practical issues related to MPR. Finally, Section VII concludes this paper.

\section{Preliminary and System Model}
\subsection{System Description}
We consider a fully conected infrastructure WLAN where \emph{N} infinitely backlogged mobile stations communicate with an access point (AP). We assume that the time axis is divided into slots and packet transmissions start only at the beginning of a slot. In addition, after each transmission, the transmitting stations have a means to discover the result of the transmission, i.e., success or failure. If the transmission fails due to collision, the colliding stations will schedule retransmissions according to a collision resolution scheme (e.g., EB). We assume that the channel has the capability to accommodate up to \emph{M} simultaneous transmissions. In other words, packets can be received correctly whenever the number of simultaneous transmissions is no larger than \emph{M}. When more than \emph{M} stations contend for the channel at the same time, collision occurs and no packet can be decoded. We refer to \emph{M} as MPR capability.

In our model, the length of a time slot is not necessarily fixed and may vary under different contexts \cite{Bianchi:00}. We refer to this variable-length slot as backoff slot hereafter. In WLANs, the length of a backoff slot depends on the contention outcome (hereafter referred to as channel status). Let $T_i$ denote the length of an idle time slot when nobody transmits; $T_c$ denote the length of a collision time slot when more than \emph{M }stations contend for the channel; and $T_s$ denote the length of a time slot due to successful transmission when the number of transmitting stations is anywhere from 1 to \emph{M}. The durations of $T_i$, $T_c$, and $T_s$ depend on the underlying WLAN configuration. For non-carrier-sensing networks such as slotted ALOHA, the stations are not aware of the channel status and the duration of all backoff slots are equal to the transmission time of a packet. That is,
\begin{equation}\label{eqn:1}
    T_{slot}=T_i=T_c=T_s=L/R
\end{equation}
where \emph{L} is the packet size and \emph{R} is the data transmission rate of a station. On the other hand, for carrier-sensing networks, stations can distinguish between various types of channel status and the durations of different types of slots may not be the same. For example, in IEEE 802.11 DCF basic access mode,
\begin{eqnarray}\label{eqn:2}
  T_i&=&\sigma \nonumber \\
  T_s&=&H+L/R+SIFS+\delta+ACK+DIFS+\delta \nonumber\\
  T_c&=&H+L/R+DIFS+\delta
\end{eqnarray}
where $\sigma$ is the time needed for a station to detect the packet transmission from any other station and is typically much smaller than $T_c$ and $T_s$; $H$ is the transmission time of PHY header and MAC header; ACK is the transmission time of an ACK packet; $\delta$ is the propagation delay; and SIFS and DIFS are the inter-frame space durations \cite{WLAN}. Similarly, in IEEE 802.11 DCF request-to-send/clear-to-send (RTS/CTS) access scheme, the slot durations are given by

\begin{eqnarray}\label{eqn:3}
T_i&=&\sigma \nonumber \\
  T_s&=&RTS+SIFS+\delta+CTS+SIFS+\delta \nonumber\\
  &&+H+L/R+SIFS+\delta+ACK+DIFS+\delta \nonumber\\
  T_c&=&RTS+DIFS+\delta
\end{eqnarray}
where $RTS$ and $CTS$ denote the transmission time of RTS and CTS packets, respectively. By allowing the durations of $T_i$, $T_c$, and $T_s$ to vary according to the underlying system, the analysis of this paper applies to a wide spectrum of various WLANs, including both non-carrier-sensing and carrier-sensing networks.

\subsection{Multiuser Detection}
This subsection briefly introduces the PHY layer MUD techniques used to decode multiple packets at the receiver. Let $x_k(n)$ denote the data symbol transmitted by user \emph{k} in symbol duration \emph{n}. If there are \emph{K} stations transmitting together, then the received signal at a receiver is given by
\begin{eqnarray}\label{eqn:4}
    \mathbf{y}(n)&=&\sum_{k=1}^K\mathbf{h}_k(n)x_k(n)+\mathbf{w}(n)\nonumber\\
    &=&\mathbf{H}(n)\mathbf{x}(n)+\mathbf{w}(n)
    \end{eqnarray}
where $\mathbf{w}(n)$ denotes the additive noise, $\mathbf{H}(n)=[\mathbf{h}_1(n), \mathbf{h}_2(n),\cdots,\mathbf{h}_K(n)]$, and $\mathbf{x}(n)=[x_1(n),\cdots,x_K(n)]^T$. In multiple antenna systems, $\mathbf{h}_k$ is the channel vector, with the $m^{th}$ element being the channel coefficient from user \emph{k} to the $m^{th}$ receive antenna.\footnote{In this paper, we assume that each station only transmits one data stream at a time.} In CDMA systems, vector $\mathbf{h}_k$ is multiplication of the spreading sequence of user \emph{k} and the channel coefficient from user \emph{k} to the AP.

The receiver attempts to obtain an estimate of the transmitted symbols $\mathbf{x}(n)$ from the received vector $\mathbf{y}(n)$. To this end, various MUD techniques have been proposed in the literature. For example, the zero-forcing (ZF) receiver is one of the most popular linear detectors. It multiplies the received vector by the pseudo-inverse of matrix $\mathbf{H}(n)$, denoted by $\mathbf{H}^\texttt{+}(n)$, and the decision statistics become
\begin{eqnarray}\label{eqn:5}
    \mathbf{r}^{ZF}(n)&=&\mathbf{H}^\texttt{+}(n)\mathbf{y}(n)\nonumber\\
    &=&\mathbf{x}(n)+\mathbf{H}^\texttt{+}(n)\mathbf{w}(n).
\end{eqnarray}
The minimum-mean-square-error (MMSE) receiver is the optimal linear detector in the sense of maximizing the signal-to-interference-and-noise ratio (SINR). The decision statistics is calculated as
\begin{equation}\label{eqn:6}
    \mathbf{r}^{MMSE}(n)=(\mathbf{H}(n)\mathbf{H}^H(n)+\eta\mathbf{I})^{-1}\mathbf{H}^H(n)\mathbf{y}(n)
\end{equation}
where $\mathbf{I}$ is the identity matrix, and $\eta$ is the variance of the additive noise. Given the decision statistics, an estimate of $x_k(n)$ can be obtained by feeding the $k^{th}$ element of $\mathbf{r}^{ZF}(n)$ or $\mathbf{r}^{MMSE}(n)$ into a quantizer.

Other MUD techniques include maximum-likelihood (ML), parallel interference cancellation (PIC), successive interference cancellation (SIC), etc. Interested readers are referred to \cite{Verdu:98} for more details.

\subsection{Exponential Backoff}
EB adaptively tunes the transmission probability of a station according to the traffic intensity of the network. It works as follows. A backlogged station sets its backoff timer by randomly choosing an integer within the range $[0,W-1]$, where \emph{W} denote the size of the contention window. The backoff timer is decreased by one following each backoff slot. The station transmits a packet in its queue once the backoff timer reaches zero. At the first transmission attempt of a packet, $W=W_0$, referred to as the minimum contention window. Each time the transmission is unsuccessful, the \emph{W} is multiplied by a backoff factor \emph{r}. That is, the contention window size $W_i=r^iW_0$ after \emph{i} successive transmission failures.

\section{Super-Linear Throughput Scaling in WLANs with MPR}
This section investigates the impact of MPR on the throughput of random-access WLANs. In particular, we prove that the maximum achievable throughput scales super-linearly with the MPR capability $M$. In practical systems, $M$ is directly related to the cost (e.g., bandwidth in CDMA systems or antenna in multi-antenna systems). Super-linear scaling of throughput implies that the achievable throughput \emph{per unit cost} increases with $M$. This provides a strong incentive to consider MPR in next-generation wireless networks.
As mentioned earlier, the transmission of stations is dictated by the underlying EB scheme. To capture the fundamentally achievable throughput of the system, the following analysis assumes that each station transmits with probability $p_t$ in an arbitrary slot, without caring how $p_t$ is achieved. The assumption will be made more rigorous in Section IV, which relates $p_t$ to EB parameters such as $r$ and $W_0$.

\subsection{Throughput of WLANs with MPR}
Define throughput to be the average number of information bits transmitted successfully per second. Let $S_N(M,p_t)$ denote the throughput of a WLAN with $N$ stations when each station transmits at probability $p_t$ and the MPR capability is $M$. Then, $S_N(M,p_t)$ can be calculated as the ratio between the average payload information bits transmitted per backoff slot to the average length of a backoff slot as follows.
\begin{equation}\label{eqn:7}
    S_N(M,p_t)=\frac{\sum_{k=1}^Mk\Pr\{X=k\}L}{P_{idle}T_i+P_{coll}T_c+P_{succ}T_s}
\end{equation}
In the above, \emph{X} is a random variable denoting the number of attempts in a slot.
\begin{equation}\label{eqn:8}
    \Pr\{X=k\}=\binom{N}{k}p_t^k(1-p_t)^{N-k}.
\end{equation}
Let
\begin{equation}\label{eqn:9}
    P_{idle}=(1-p_t)^N
\end{equation}
be the probability that a backoff slot is idle;
\begin{equation}\label{eqn:10}
    P_{succ}=\sum_{k=1}^M\Pr\{X=k\}=\sum_{k=1}^M\binom{N}{k}p_t^k(1-p_t)^{N-k}
\end{equation}
be the probability that a backoff slot is busy due to successful packet transmissions; and
\begin{equation}\label{eqn:11}
    P_{coll}=\sum_{k=M+1}^N\Pr\{X=k\}=\sum_{k=M+1}^N\binom{N}{k}p_t^k(1-p_t)^{N-k}
\end{equation}
be the probability that a backoff slot is busy due to collision of packets.

The throughput of non-carrier-sensing networks such as slotted ALOHA can be obtained by substituting (\ref{eqn:1}) into (\ref{eqn:7}), which leads to following expression:
\begin{eqnarray}\label{eqn:12}
    S_N(M,p_t)&=&\frac{\sum_{k=1}^Mk\Pr\{X=k\}L}{T_{slot}}\nonumber\\
    &=&R\sum_{k=1}^Mk\binom{N}{k}p_t^k(1-p_t)^{N-k}
\end{eqnarray}
Similarly, the throughput of carrier-sensing networks, such as IEEE 802.11 DCF basic-access mode and RTS/CTS access mode, can be obtained by substituting (\ref{eqn:2}) and (\ref{eqn:3}) into (\ref{eqn:7}) respectively.

We now derive the asymptotic throughput when the population size \emph{N} approaches infinity. In this case, we assume that (i) the system has a non-zero asymptotic throughput; and (ii) the number of attempts in a backoff slot is approximated by a Poisson distribution with an average attempt rate
$\lambda=Np_t$ \cite[pp. 258]{DeGroot}. Both of these assumptions are valid under an appropriate EB scheme, which will be elaborated in Section IV. Let $S_{\infty}(M,\lambda)$ be the asymptotic throughput when MPR capability is \emph{M} and average attempt rate is $\lambda$. Then, we derive from (\ref{eqn:7}) that

\begin{eqnarray}\label{eqn:13}
    S_{\infty}(M,\lambda)&=&\lim_{N\rightarrow\infty}S_N\nonumber\\
    &=&\frac{L\sum_{k=1}^Mk\Pr\{X=k\}}{P_{idle}T_i+P_{coll}T_c+P_{succ}T_s}\nonumber\\
    &=&\frac{L\sum_{k=1}^Mk\frac{\lambda^k}{k!}e^{-\lambda}}{P_{idle}T_i+P_{coll}T_c+P_{succ}T_s}\nonumber\\
    &=&\frac{L\lambda\sum_{k=0}^{M-1}\frac{\lambda^k}{k!}e^{-\lambda}}{P_{idle}T_i+P_{coll}T_c+P_{succ}T_s}\nonumber\\
    &=&\frac{L\lambda\Pr\{X\leq M-1\}}{P_{idle}T_i+P_{coll}T_c+P_{succ}T_s}
\end{eqnarray}
where the third equality is due to the Poisson approximation. In particular, when $T_{slot}=T_i=T_c=T_s=L/R$,
\begin{eqnarray}\label{eqn:14}
  S_{\infty}(M,\lambda)&=&R\sum_{k=0}^{M-1}\frac{\lambda^{k+1}}{k!}e^{-\lambda}\nonumber\\
   &=& R\lambda\Pr\{X \leq M-1\}
\end{eqnarray}

\subsection{Super-Linear Throughput Scaling}
Having derived the throughput expressions for both finite-population and infinite-population models, we now address the question: how does throughput scale as \emph{M} increases. In particular, we are interested in the behavior of the maximum throughput when the channel has a MPR capability of \emph{M}. This directly relates to the channel-access efficiency that is achievable in MPR networks.

Given \emph{M}, the maximum throughput can be achieved by optimizing the transmission probability $p_t$ (or equivalently $\lambda$ in the infinite-population model). The optimal transmission probability can in turn be obtained by adjusting the backoff factor \emph{r} in practical WLANs, as will be discussed in Section IV. Let $S_N^*(M)=S_N(M,p_t^*(M))$ and $S_{\infty}^*(M)=S_{\infty}(M,\lambda^*(M))$ denote the maximum achievable throughputs, where $p_t^*(M)$ and $\lambda^*(M)$ denote the optimal $p_t$ and $\lambda$ when the MPR capability is \emph{M}, respectively.  In Theorem 1, we prove that the throughput scales super-linearly with \emph{M} in non-carrier-sensing network with infinite population. In other words, $S_{\infty}^*(M)/M$ is an increasing function of \emph{M}. In Theorem 2, we further prove that $S_{\infty}^*(M)/MR$ approaches 1 when $M\rightarrow\infty$. This implies that the throughput penalty due to distributed random access diminishes when \emph{M} is very large. In Theorem 3 in Appendix I, we prove that the same super-linearity holds for WLANs with finite population.

\begin{theorem} \label{theorem:1}
\emph{(Super-Linearity)} $S_{\infty}^*(M)/M$ is an increasing function of \emph{M}.

It is obvious that at the optimal $\lambda^*(M)$

\begin{eqnarray}\label{eqn:15}
    \frac{\partial S_{\infty}(M,\lambda)}{\partial\lambda}\bigg|_{\lambda=\lambda^*(M)}&=&R\sum_{k=0}^{M-1}\frac{(k+1)(\lambda^*(M))^k}{k!}e^{-\lambda^*(M)}\nonumber\\
    &&-R\sum_{k=0}^{M-1}\frac{(\lambda^*(M))^{(k+1)}}{k!}e^{-\lambda^*(M)}\nonumber\\
    &=&0
\end{eqnarray}
Consequently,
\begin{equation}\label{eqn:16}
\sum_{k=0}^{M-1}\frac{(\lambda^*(M))^k}{k!}e^{-\lambda^*(M)}=\frac{(\lambda^*(M))^M}{(M-1)!}e^{-\lambda^*(M)},
\end{equation}
or
\begin{equation}\label{eqn:17}
\Pr\{X \leq M-1\}\bigg|_{\lambda=\lambda^*(M)}=M\Pr\{X=M\}\bigg|_{\lambda=\lambda^*(M)}.
\end{equation}

To prove Theorem 1, we show that $S_{\infty}^*(M+1)/(M+1) \geq S_{\infty}^*(M)/M$ for all \emph{M} in the following.

\begin{eqnarray}\label{eqn:18}
    &&S_{\infty}^*(M+1)=S_{\infty}(M+1,\lambda^*(M+1)) \nonumber\\
    &\geq& S_{\infty}(M+1,\lambda^*(M))\nonumber\\
    &=& R\sum_{k=0}^{M-1}\frac{\lambda^*(M)^{k+1}}{k!}e^{-\lambda^*(M)}\nonumber\\
    &&+R\frac{\lambda^*(M)^{M+1}}{M!}e^{-\lambda^*(M)}\nonumber\\
    &=&S_{\infty}(M,\lambda^*(M))+R\lambda^*(M)\Pr\{X=M\}\bigg|_{\lambda=\lambda^*(M)}\nonumber\\
    &=&\frac{M+1}{M}S_{\infty}^*(M)
\end{eqnarray}
where the last equality is due to (\ref{eqn:14}) and (\ref{eqn:17}). Therefore, we have
\begin{eqnarray}
    \frac{S_{\infty}(M+1)}{M+1} \geq \frac{S_{\infty}(M)}{M}\; \forall M \nonumber
\end{eqnarray}
\begin{flushright}
$\square$
\end{flushright}
\end{theorem}
It is obvious that in a WLAN with MPR capability of $M$, the maximum possible throughput is $MR$ when there exists a perfect scheduling. In practical random-access WLANs, the actual throughput is always smaller than $MR$, due to the throughput penalty resulting from packet collisions and idle slots. For example, the maximum throughput is well known to be $Re^{-1}$  when $M=1$. Theorem 2 proves that the throughput penalty diminishes as $M$ becomes large. That is, the maximum throughput approaches $MR$ even though the channel access is based on random contentions.

\begin{theorem} \label{theorem:2}
\emph{(Asymptotic channel-access efficiency)} $\lim_{M\rightarrow\infty}S_{\infty}^*(M)\big/MR=1$.

Before proving Theorem 2, we present the following two lemmas.
\begin{lemma}
(a) $\lim_{M\rightarrow\infty}S_\infty(M)\big/\lambda R=1$ for any attempt rate $\lambda<M$; (b) $\lim_{M\rightarrow\infty}S_\infty(M)\big/\lambda R=0$ for any attempt rate $\lambda>M$; (c) $\lim_{M\rightarrow\infty}S_\infty(M)\big/\lambda R=0.5$ for attempt rate $\lambda=M$.
\end{lemma}

\emph{Proof of Lemma 1(a):}

\begin{eqnarray}\label{eqn:19}
S_{\infty}(M,\lambda)&=& R\lambda\Pr\{X \leq M-1\}\nonumber\\
&=& R\lambda\big(1-\sum_{k=M}^{\infty}\frac{\lambda^k}{k!}e^{-\lambda}\big)\nonumber\\
&\geq&R\lambda\big(1-z^{-M}\sum_{k=M}^{\infty}\frac{(\lambda z)^k}{k!}e^{-\lambda}\big)\nonumber\\
&\geq&R\lambda\big(1-z^{-M}e^{\lambda(z-1)}\big)\;  \forall z>1
\end{eqnarray}
Let $f(z)=R\lambda\big(1-z^{-M}e^{\lambda (z-1)}\big)$ be the lower bound of $S_{\infty}(M)$. By solving
\begin{equation}\label{eqn:20}
    \frac{\partial f(x)}{\partial z}=R\lambda\big(Mz^{-M-1}e^{\lambda(z-1)}-\lambda z^{-M}e^{\lambda(z-1)}\big)=0
\end{equation}
it can be easily found that $z^*=M\big/\lambda$ maximizes $f(z)$ and
\begin{equation}\label{eqn:21}
    \frac{f(z^*)}{\lambda R}=1-\bigg(\frac{\lambda}{M}\bigg)^Me^{M(1-\frac{\lambda}{M})}
    \end{equation}
Since $z^*>1$, $\lambda<M$. Let $\lambda=cM$ where $c<1$. eqn. (\ref{eqn:21}) can be written as

\begin{equation}\label{eqn:22}
\frac{f(z^*)}{\lambda R}=1-\big(ce^{1-c}\big)^M
\end{equation}

It is obvious that
\begin{equation}\label{eqn:23}
ce^{1-c}<1 \;\forall c\neq 1.
\end{equation}
Therefore,
\begin{eqnarray}\label{eqn:24}
  \lim_{M\rightarrow\infty}\frac{S_{\infty}(M,\lambda)}{\lambda R}&\geq&\lim_{M\rightarrow\infty}\frac{f^*(z)}{\lambda R}\nonumber\\
  &=&\lim_{M\rightarrow\infty}\bigg(1-\big(ce^{1-c}\big)^M\bigg)\nonumber\\
  &=&1.
\end{eqnarray}
On the other hand, the first equality of (\ref{eqn:19}) implies
\begin{equation}\label{eqn:25}
    \frac{S_\infty (M,\lambda)}{\lambda R}\leq 1.
\end{equation}
Combining (\ref{eqn:24}) and (\ref{eqn:25}), we have
\begin{equation}\label{eqn:26}
    \lim_{M\rightarrow\infty}\frac{S_\infty (M,\lambda)}{\lambda R}= 1 \; \forall \lambda<M,
\end{equation}
and Lemma 1(a) follows.
\begin{flushright}
$\Box$
\end{flushright}

\emph{Proof of Lemma 1(b)}:
\begin{eqnarray}\label{eqn:B1}
    S_\infty(M)&=&R\lambda\Pr\{X\leq M-1\}=R\lambda\sum_{k=0}^{M-1}\frac{\lambda^k}{k!}e^{-\lambda}\nonumber\\
    &\leq&R\lambda z^{-M}\sum_{k=0}^{M-1}\frac{(\lambda z)^k}{k!}e^{-\lambda}\nonumber\\
    &\leq&R\lambda z^{-M}\sum_{k=0}^\infty\frac{(\lambda z)^k}{k!}e^{-\lambda}\nonumber\\
    &=&R\lambda z^{-M}e^{\lambda(z-1)}\;\forall z<1.
\end{eqnarray}
Let $g(z)=R\lambda z^{-M}e^{\lambda(z-1)}$ be the upper bound of $S_\infty(M)$. By solving
\begin{equation}\label{eqn:B2}
    \frac{\partial g(z)}{\partial z}=R\lambda\big(-Mz^{-M-1}e^{\lambda(z-1)}+\lambda z^{-M}e^{\lambda(z-1)}\big)=0
\end{equation}
it can be easily found that $z^*=M/\lambda$ minimizes $g(z)$ and
\begin{equation}\label{eqn:B3}
    \frac{g(z^*)}{R\lambda}=\bigg(\frac{\lambda}{M}\bigg)^Me^{M(1-\frac{\lambda}{M})}.
\end{equation}
Since $z^*<1$, $\lambda>M$. Let $\lambda=cM$ where $c>1$. eqn. (\ref{eqn:B3}) can be written as
\begin{equation}\label{eqn:B4}
    \frac{g(z^*)}{R\lambda}=\bigg(ce^{1-c}\bigg)^M.
\end{equation}
Due to eqn. (\ref{eqn:23})
\begin{eqnarray}\label{eqn:B5}
    \lim_{M\rightarrow\infty}\frac{S_\infty(M)}{R\lambda}&\leq& \lim_{M\rightarrow\infty}\frac{g^*(z)}{R\lambda}\nonumber\\
    &=&\lim_{M\rightarrow\infty}\big(ce^{1-c}\big)^M=0
\end{eqnarray}
On the other hand, it is obvious that
\begin{equation}\label{eqn:B6}
    \frac{S_\infty(M)}{R\lambda}\geq0
\end{equation}
Combining (\ref{eqn:B5}) and (\ref{eqn:B6}), we have
\begin{equation}\label{eqn:B7}
    \lim_{M\rightarrow\infty}\frac{S_\infty(M)}{R\lambda}=0\;\forall\lambda>M,
\end{equation}
and Lemma 1(b) follows.

\emph{Proof of Lemma 1(c)}:

To prove Lemma 1(c), we note that the median of Poisson distribution is bounded as follows \cite{Choi:94}-\cite{Chen:86}:
\begin{equation}\label{eqn:B8}
    \lambda-\log2\leq median \leq \lambda+1/3.
\end{equation}
When $\lambda=M$ and $M\rightarrow\infty$, the median approaches $M$. According to the first equality of (\ref{eqn:14}),
\begin{eqnarray}\label{eqn:B9}
    \lim_{M\rightarrow\infty}\frac{S_\infty(M)}{R\lambda}&=&\lim_{M\rightarrow\infty}\Pr\{X\leq M-1\} \nonumber\\
    &\approx &\lim_{M\rightarrow\infty}\Pr\{X\leq M\} =0.5
\end{eqnarray}
\begin{flushright}
$\Box$
\end{flushright}

\begin{lemma}
The optimal attempt rate $\lambda^*(M)<M$ and $\lim_{M\rightarrow\infty}\lambda^*(M)\big/M=1$.
\end{lemma}
\emph{Proof of Lemma 2:} The mode of Poisson distribution is equal to $\lfloor\lambda\rfloor$, where $\lfloor\cdot\rfloor$ denotes the largest integer that is smaller than or equal to the argument. When $\lambda\geq M$,
\begin{equation}\label{eqn:27}
    \Pr\{X=M\}>\Pr\{X=i\}\;\forall 0\leq i\leq M-1,
\end{equation}
which conflicts with eqn. (\ref{eqn:17}). Therefore, the optimal attempt rate
\begin{equation}\label{eqn:28}
    \lambda^*(M)<M.
\end{equation}
Combining (\ref{eqn:14}), (\ref{eqn:17}), (\ref{eqn:28}) and Lemma 1, we have
\begin{equation}\label{eqn:29}
    \lim_{M\rightarrow\infty}M\Pr\{X=M\}\big|_{\lambda=\lambda^*(M)}=1.
\end{equation}
Let $\lambda^*=cM$ where $c<1$. eqn. (\ref{eqn:29}) can be written as
\begin{equation}\label{eqn:30}
    \lim_{M\rightarrow\infty}\frac{(cM)^M}{(M-1)!}e^{-cM}=1.
\end{equation}
and
\begin{eqnarray}\label{eqn:31}
  c&=& \lim_{M\rightarrow\infty}\frac{\big((M-1)!\big)^{1/M}}{M}e^c\nonumber \\
   &\approx&\lim_{M\rightarrow\infty} \frac{(M!)^{1/M}}{M}e^c \nonumber\\
  &=& e^{-(1-c)}
\end{eqnarray}
where the last equality is due to the Stirling's formula \cite{Feller:68}.
Solving eqn. (\ref{eqn:31}), we have
\begin{equation}
    \lim_{M\rightarrow\infty}\frac{\lambda^*}{M}=\lim_{M\rightarrow\infty}c=1
\end{equation}
\begin{flushright}
$\Box$
\end{flushright}

\emph{Proof of Theorem 2}: From Lemma 1 and Lemma 2, it is obvious that $\lim_{M\rightarrow\infty}S_{\infty}^*(M)\big/MR=1$.
\begin{flushright}
$\Box$
\end{flushright}
\end{theorem}

\begin{figure}
\centering
\includegraphics[width=0.5\textwidth]{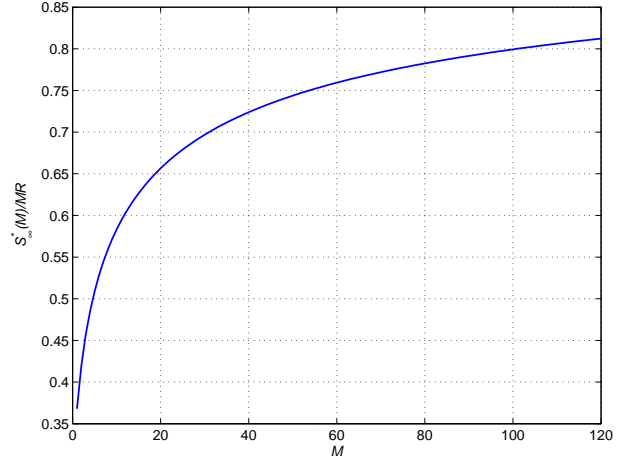}
\caption{Super-linear scalability of the throughput of non-carrier-sensing slotted ALOHA networks}\label{fig:1}
\end{figure}

The above results are illustrated in Fig. \ref{fig:1}, where $S_\infty^*(M)\big/MR$ is plotted as a function of $M$ in non-carrier-sensing slotted ALOHA systems.

\begin{theorem}
\emph{(Super-linearity with finite population)} $S_N^*(M+1)\big/M+1\geq S_N^*(M)\big/M$ for all $M<N$.
\end{theorem}

\emph{Proof of Theorem 3}: See Appendix I.

\begin{figure}
\centering
\includegraphics[width=0.5\textwidth]{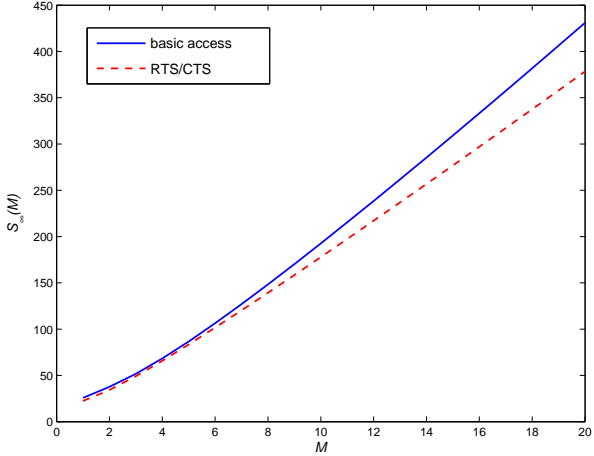}
\caption{Optimal throughput of carrier-sensing networks}\label{fig:2}
\end{figure}

\begin{figure}
\centering
\includegraphics[width=0.5\textwidth]{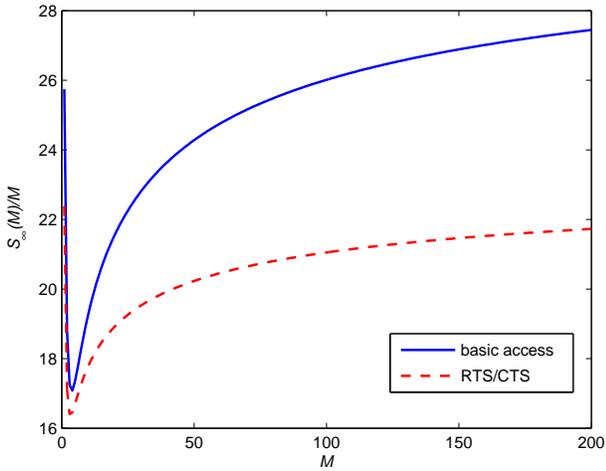}
\caption{Super-linear scalability of the throughput of carrier-sensing networks}\label{fig:3}
\end{figure}

In Theorem 1-3, super-linearity is proved assuming the network is non-carrier-sensing. In Fig. \ref{fig:2} and Fig. \ref{fig:3}, the optimal throughput $S_\infty^*(M)$ and $S_\infty^*(M)\big/M$ are plotted for carrier-sensing networks, respectively, with system parameters listed in Table I. The figures show that system throughput is greatly enhanced due to the MPR enhancement in the PHY layer. Moreover, the super-linear throughput scaling holds for carrier-sensing networks when \emph{M} is relatively large.
\begin{table}
\caption{System Parameters Used in Carrier-Sensing Networks (Adopted from IEEE 802.11g)}
\centering
\begin{tabular}{|c|c|}
  \hline
  Packet payload & 8184 bits \\
  \hline
  MAC header & 272 bits \\
  \hline
  PHY overhead & 26 $\mu s$ \\
  \hline
  ACK & 112 bits + PHY overhead \\
  \hline
  RTS & 160 bits + PHY overhead \\
  \hline
  CTS & 112 bits + PHY overhead \\
  \hline
  Basic rate & 6 Mbps \\
  \hline
  Data rate & 54 Mbps \\
  \hline
  Slot time $\sigma$ & 9 $\mu s$ \\
  \hline
  SIFS & 10 $\mu s$ \\
  \hline
\end{tabular}
\end{table}

\section{Impact of MPR on EB in WLAN MAC}
In this section, we study the characteristic behavior of WLAN MAC and EB when the channel has MPR capability. We first establish the relationship between transmission probability $p_t$ (or $\lambda$) and EB parameters including backoff factor \emph{r} and minimum contention window $W_0$. Based on the analysis, we will then study how the optimal backoff strategy changes with the MPR capability $M$.

\subsection{Transmission Probability}
\begin{figure}
\centering
\includegraphics[width=0.5\textwidth]{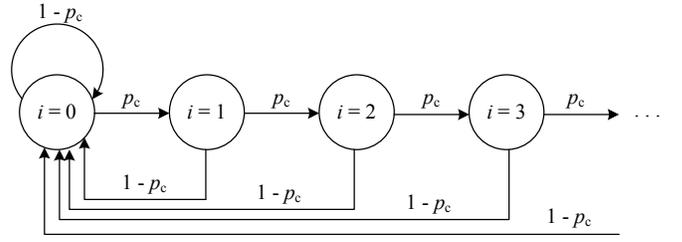}
\caption{Markov chain model for the backoff stage}\label{fig:4}
\end{figure}

We use an infinite-state Markov chain, as shown in Fig. \ref{fig:4}, to model of operation of EB with no retry limit. The reason for the lack of a retry limit is that it is theoretically more interesting to look at the limiting case when the retry limit is infinitely large. Having said this, we note that the analysis in our paper can be easily extended to the case where there is a retry limit. The state in the Markov chain in Fig. \ref{fig:4} is the backoff stage, which is also equal to the number of retransmissions experienced by the station. As mentioned in Section II, the contention window size is $W_i=r^iW_0$ when a station is in state $i$. In the figure, $p_c$ denotes the conditional collision probability, which is the probability of a collision seen by a packet being transmitted on the channel. Note that $p_c$ depends on the transmission probabilities of stations other than the transmitting one. In our model, $p_c$  is assumed to be independent of the backoff stage of the transmitting station. In our numerical results, we show that the analytical results obtained under this assumption are very accurate when $N$ is reasonably large.

\begin{figure}
\centering
\includegraphics[width=0.5\textwidth]{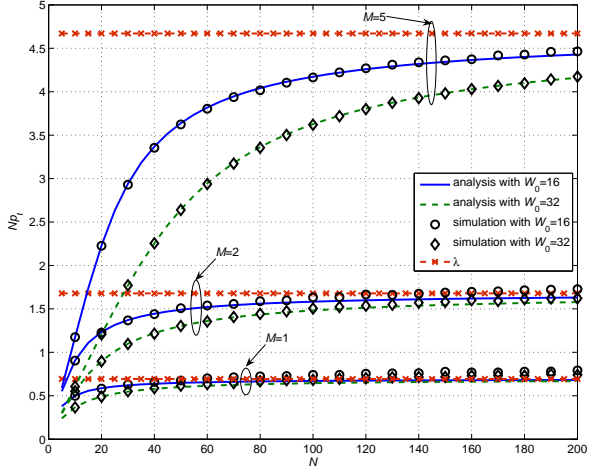}
\caption{Plots of $Np_t$ versus $N$ when $r=2$; lines are analytical results calculated from (\ref{eqn:2}) and (\ref{eqn:3}), markers are simulation results.}\label{fig:5}
\end{figure}

With EB, transmission probability $p_t$ is equal to the probability that the backoff timer of a station reaches zero in a slot. Note that the Markov process of MPR networks is similar to the ones in \cite{Bianchi:00}, \cite{Kwak:05}, except that the conditional collision probability $p_c$ is different for $M>1$. Therefore, eqn. (\ref{eqn:32}) can be derived in a similar way as \cite{Bianchi:00}, \cite{Kwak:05}:
\begin{equation}\label{eqn:32}
    p_t=\frac{2(1-rp_c)}{W_0(1-p_c)+1-rp_c}
\end{equation}
where $rp_c<1$ is a necessary condition for the steady state to be reachable. The detailed derivation of (\ref{eqn:32}) is omitted due to page limit. Interested readers are referred to \cite{Bianchi:00}, \cite{Kwak:05}. Likewise, the conditional collision probability $p_c$ is equal to the probability that there are $M$ or more stations out of the remaining $N-1$ stations contending for the channel. We thus have the following relationship:
\begin{equation}\label{eqn:33}
    p_c=1-\sum_{k=0}^{M-1}\binom{N-1}{k}p_t^k(1-p_t)^{N-k-1}.
\end{equation}
It can be easily shown that $p_t$ is a decreasing function of $p_c$ for any $r>1$ in (\ref{eqn:32}). Meanwhile, $p_c$ is an increasing function of $p_t$ in (\ref{eqn:33}). Therefore, the curves determined by (\ref{eqn:32}) and (\ref{eqn:33}) have a unique intersection corresponding to the root of the nonlinear system. By solving the nonlinear system (\ref{eqn:32})-(\ref{eqn:33}) numerically for different $N$, we plot the analytical results of $Np_t$ in Fig. \ref{fig:5}. In the figures, BEB is adopted. That is, $r=2$. The minimum contention window size $W_0=16$ or 32. To validate the analysis, the simulation results are plotted as markers in the figures. In the simulation, the data are collected by running 5,000,000 rounds after 1,000,000 rounds of warm up.
From the figures, we can see that the analytical results match the simulations very well. Moreover, it shows that $Np_t$ converges to a constant quantity when $N$ becomes large. This is a basic assumption in the previous section when we calculated the asymptotic throughput. The constant quantity that $Np_t$ converges to can be calculated as follows.

For large $N$, the number of attempts in a slot can be modeled as a Poisson process \cite[pp. 258]{DeGroot}. That is,
\begin{equation}\label{eqn:34}
    \Pr\{X=k\}=\frac{\lambda^k}{k!}e^{-\lambda}
\end{equation}
where
\begin{equation}\label{eqn:35}
    \lambda=\lim_{N\rightarrow\infty}Np_t.
\end{equation}
The conditional collision probability in this limiting case is given by
\begin{equation}\label{eqn:36}
    \lim_{N\rightarrow\infty}p_c=\Pr\{X\geq M\}=1-\sum_{k=0}^{M-1}\frac{\lambda^k}{k!}e^{-\lambda}.
\end{equation}

When the system is steady, the total attempt rate $\lambda=\lim_{N\rightarrow\infty}Np_t$ should be finite. Therefore,
\begin{equation}\label{eqn:37}
    \lim_{N\rightarrow\infty}p_t=\lim_{N\rightarrow\infty}\frac{2(1-rp_c)}{W_0(1-p_c)+1-rp_c}=0,
\end{equation}
which implies
\begin{equation}\label{eqn:38}
    \lim_{N\rightarrow\infty}p_c=\frac{1}{r}.
\end{equation}
Combining (\ref{eqn:36}) and (\ref{eqn:38}), we get the following equation
\begin{equation}\label{eqn:39}
    \sum_{k=0}^{M-1}\frac{\lambda^k}{k!}e^{-\lambda}=1-\frac{1}{r}.
\end{equation}
$\lambda$ can be calculated numerically from (\ref{eqn:39}) given $M$ and $r$. Fig. \ref{fig:5} shows that $Np_t$ calculated from (\ref{eqn:32}) and (\ref{eqn:33}) does converge to $\lambda$ when $N$ is large.

Note that the relationship between $p_t$, $\lambda$, and EB established above do not depend on the duration of the underlying backoff slots, and therefore can be applied in both non-carrier-sensing and carrier-sensing networks.

\begin{figure}
\centering
\includegraphics[width=0.5\textwidth]{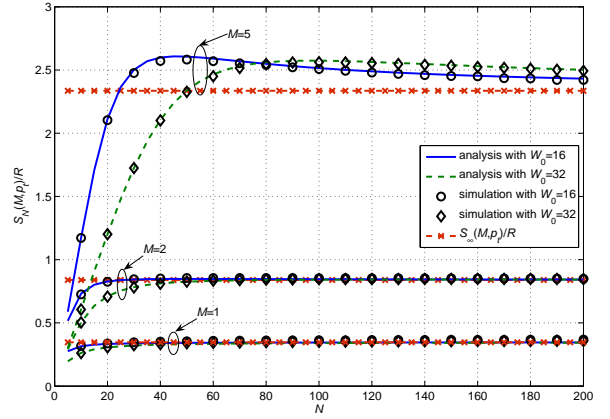}
\caption{Normalized throughput of non-carrier-sensing slotted ALOHA networks when $r=2$}\label{fig:6}
\end{figure}

\begin{figure}
\centering
\includegraphics[width=0.5\textwidth]{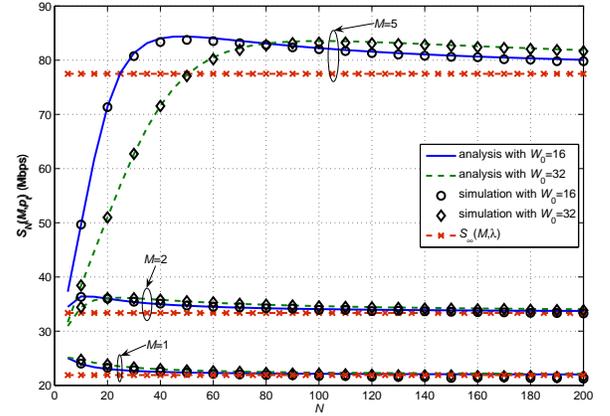}
\caption{Throughput of carrier-sensing basic-access networks when $r=2$}\label{fig:7}
\end{figure}

Before leaving this sub-section, we validate another assumption adopted in Section III. That is, EB guarantees a non-zero throughput when $N$ approaches infinity. To this end, the throughput of slotted ALOHA is plotted as a function of $N$ in Fig. \ref{fig:6} when BEB is adopted. It can be seen that the throughputs with the same $M$ converge to the same constant as $N$ increases, regardless of the minimum contention window $W_0$. Similar phenomenon can also be observed in carrier-sensing networks, as illustrated in Fig. \ref{fig:7}, where the throughput of IEEE 802.11 WLAN with basic access mode is plotted with detailed system parameters listed in Table I. The asymptotic throughput when $N$ is very large depends only on the MPR capability $M$ and the backoff factor $r$.

\subsection {Optimal Backoff Factor}
In Section III, we have investigated the maximum network throughput that is achieved by optimal transmission probability $p_t^*(M)$ and $\lambda^*(M)$. The previous sub-section shows that transmission probability is a function of backoff factor $r$. Mathematically, the optimal $r$ that maximizes throughput can be obtained by solving the equation $\partial S(M)\big/\partial r=0$.

\begin{figure}
\centering
\includegraphics[width=0.5\textwidth]{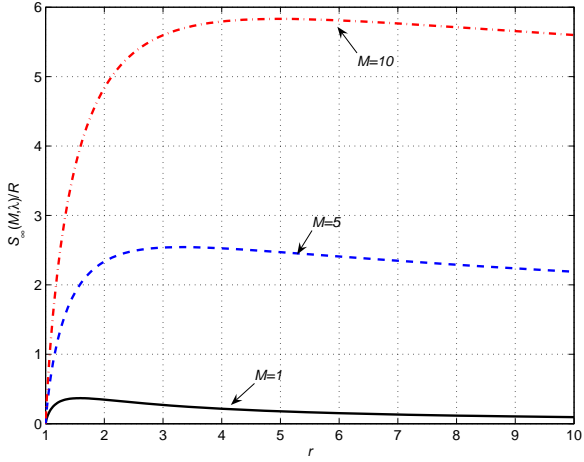}
\caption{Throughput versus $r$ for non-carrier-sensing slotted ALOHA networks}\label{fig:8}
\end{figure}

\begin{figure}
\centering
\includegraphics[width=0.5\textwidth]{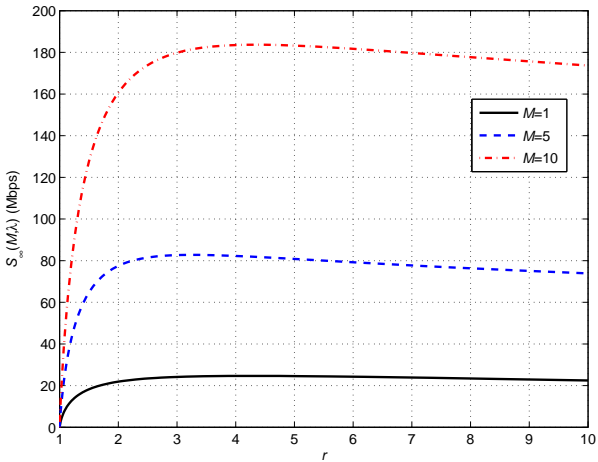}
\caption{Throughput versus $r$ for carrier-sensing networks with basic-access mode}\label{fig:9}
\end{figure}

In this section, we investigate how the optimal backoff factor $r$ changes with the MPR capability $M$.  In Fig. \ref{fig:8} and Fig. \ref{fig:9}, we plot the throughput as a function of $r$ for both non-carrier sensing networks and carrier-sensing networks in basic-access mode. From the figure, we can see that the optimal $r$ that maximizes throughput increases with $M$ for moderate to large $M$. This observation can be intuitively explained for non-carrier-sensing networks by (\ref{eqn:14}), (\ref{eqn:39}), and Lemma 1 as follows. Eqns. (\ref{eqn:14}) and (\ref{eqn:39}) indicate that
\begin{equation}\label{eqn:40}
    \frac{S_{\infty}(M,\lambda)}{R\lambda}=\Pr\{X\leq M-1\}=1-\frac{1}{r}.
\end{equation}
As Lemma 1 indicates, when $M$ is large, $S_{\infty}(M,\lambda)\big/R\lambda$ increases with $M$ and eventually approaches 1. Consequently, $r$ increases with $M$.

\begin{figure}
\centering
\includegraphics[width=0.5\textwidth]{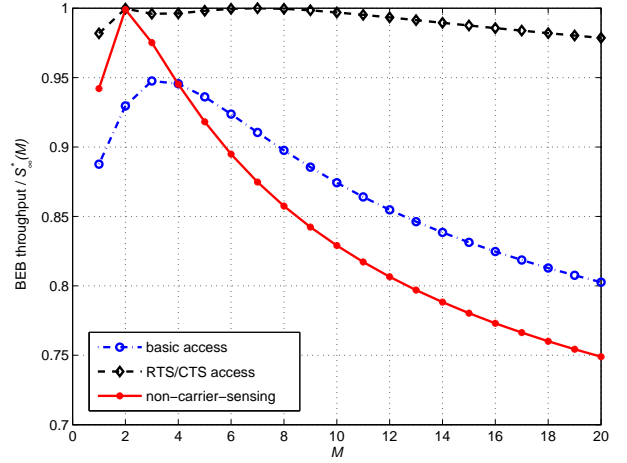}
\caption{Ratio of BEB throughput to the maximal throughput versus $M$}\label{fig:10}
\end{figure}

\begin{figure}
\centering
\includegraphics[width=0.5\textwidth]{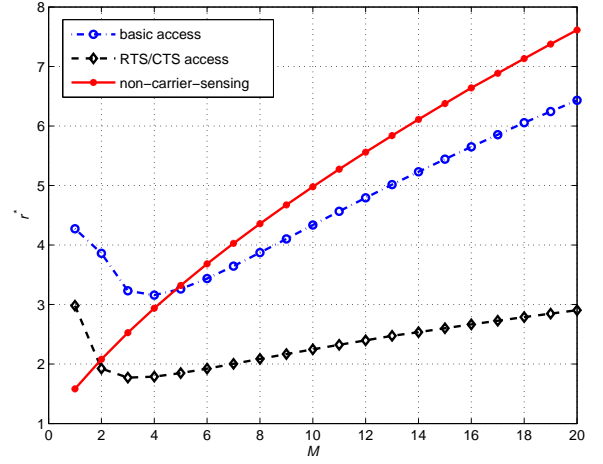}
\caption{Optimal $r$ versus $M$}\label{fig:11}
\end{figure}

As the figures show, the throughput decreases sharply when $r$ moves from the optimal $r^*$ to 1. On the other hand, it is much less sensitive to $r$ when $r$ is larger than the $r^*$. Therefore, in order to avoid dramatic throughput degradation, it is not wise to operate $r$ in the region between 1 and $r^*$. Note that when $M$ is large, $r^*$ is larger than 2. This implies that the widely used BEB might be far from optimal in MPR WLANs. To further see how well BEB works, we plot the ratio of the throughput obtained by BEB to the maximum achievable throughput in Fig. \ref{fig:10}. The optimal $r$ that achieves the maximum throughput is plotted versus $M$ in Fig. \ref{fig:11}. In the figures, we can see that BEB only achieves a small fraction of the maximum achievable throughput when $M$ is large in non-carrier-sensing and IEEE 802.11 basic-access mode. For example, when $M=10$ BEB only achieves about 80 percent of the maximum throughput in non-carrier-sensing networks. In RTS/CTS mode, in contrast, the performance of BEB is close to optimal for a large range of $M$. Therefore, we argue from an engineering point of view that BEB (i.e., $r=2$) is a good choice for RTS/CTS access scheme, while on the other hand tuning $r$ to the optimal is important for non-carrier-sensing and basic-access schemes.

Having demonstrated the significant capacity improvement that MPR brings to WLANs, we are highly motivated to present practical protocols to implement MPR in the widely used IEEE 802.11 WiFi. In particular, we will propose protocols that consist of both MAC-layer mechanisms and PHY-layer signal processing schemes in the next Section.

\section{MPR Protocol for IEEE 802.11 WLAN}
In this section, we present a MPR protocol for IEEE 802.11 WLAN with RTS/CTS mechanism. The proposed protocol requires minimum amendment at mobile stations, and hence will be easy to implement in practical systems.
Throughout this section, we assume that the MPR capability is brought by the multiple antennas mounted at the access point (AP). This assumption complies with the hardware request of the latest MIMO-based WLAN standards. However, the proposed MAC-PHY protocol can be easily extended to CDMA networks, as the received signal structures in multi-antenna and CDMA systems are almost the same (refer to Section II-C).

\subsection{MAC Protocol Design}
\begin{figure}
\centering
\includegraphics[width=0.5\textwidth]{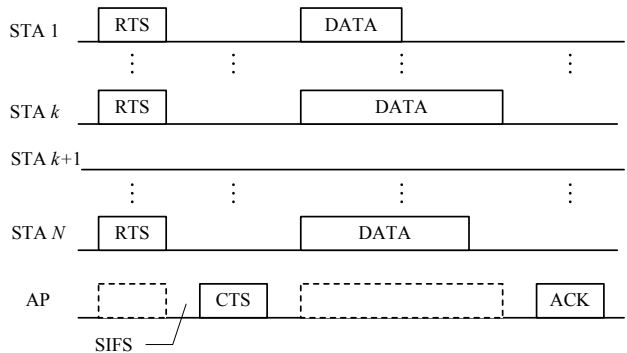}
\caption{Time line example for the MPR MAC}\label{fig:12}
\end{figure}

The MAC protocol closely follows the IEEE 802.11 RTS/CTS access mechanism, as illustrated in Fig. \ref{fig:12}. A station with a packet to transmit first sends an RTS frame to the AP. In our MPR MAC model, when multiple stations transmit RTS frames at the same time, the AP can successfully detect all the RTS frames if and only if the number of RTSs is no larger than $M$. When the number of transmitting stations exceeds $M$, collisions occur and the AP cannot decode any of the RTSs. The stations will retransmit their RTS frames after a backoff time period according to the original IEEE 802.11 protocol. When the AP detects the RTSs successfully, it responds, after a SIFS period, with a CTS frame that grants transmission permissions to all the requesting stations. Then the transmitting stations will start transmitting DATA frames after a SIFS, and the AP will acknowledge the reception of the DATA frames by an ACK frame.

The formats of the RTS and Data frames are the same as those defined in 802.11, while the CTS and ACK frames have been modified to accommodate multiple transmitting stations for MPR. In particular, there are $M$ receiver address fields in the CTS and ACK frames to identify up to $M$ intended recipients.

As described above, our MPR MAC is very similar to the original IEEE 802.11 MAC. In fact, to maintain this similarity in the MAC layer, the challenge is pushed down to the physical layer. For example, in the proposed MPR MAC, the AP is responsible to decode all the RTSs transmitted simultaneously. However, due to the random-access nature WLAN, the AP has no priori knowledge of who the senders are and the channel state information (CSI) on the corresponding links.  This imposes a major challenge on the PHY layer, as the MUD techniques introduced in Section II, such as ZF and MMSE cannot be directly applied. To tackle these problems, we introduce the physical layer techniques in next subsection.

\subsection{PHY-layer Signal Processing Mechanism}
In this subsection, we propose a PHY mechanism to implement MPR in IEEE 802.11. The basic idea is as follows. RTS packets are typically transmitted at a lower data rate than the data packets in IEEE 802.11. This setting is particularly suitable for blind detection schemes which can separate the RTS packets without knowing the prior knowledge of the senders' identities and CSI \cite{Talwar:96}-\cite{Papadias:97}. Upon successfully decoding the RTS packets, the AP can then identify the senders of the packets. Training sequences, to be transmitted in the preamble of the data packets, are then allocated to these users to facilitate channel estimation during the data transmission phase. Since the multiple stations transmit their data packets at the same time, their training sequences should be mutually orthogonal. In our system, no more than $M$ simultaneous transmissions are allowed. Therefore, a total of $M$ orthogonal sequences are required to be predefined and made known to all stations. The sequence allocation decision is sent to the users via the CTS packet.

During the data transmission phase, CSI is estimated from the orthogonal training sequences that are transmitted in the preamble of the data packets. With the estimated CSI, various MUD techniques can be applied to separate the multiple data packets at the AP. Using coherent detection, data packets can be transmitted at a much higher rate than the RTS packets without involving excessive computational complexity.

As MUD techniques have been introduced in Section II, we focus on the blind separation of RTS packets in this subsection. Assume that there are $K$ stations transmitting RTS packets together. Then, the received signal in symbol duration $n$ is given by (\ref{eqn:4}), where the $(m, k)^{th}$ element of $\mathbf{H}$ denotes the channel coefficient from user $k$ to the $m^{th}$ antenna at the AP. Assuming that the channel is constant over an RTS packet, which is composed of $N$ symbol periods, we obtain the following block formulation of the data
\begin{equation}\label{eqn:41}
    \mathbf{Y}=\mathbf{HX}+\mathbf{W}
\end{equation}
where $\mathbf{Y}=[\mathbf{y}(1),\mathbf{y}(2),\cdots,\mathbf{y}(N)]$, $\mathbf{X}=[\mathbf{x}(1),\mathbf{x}(2),\cdots,\mathbf{x}(N)]$, and $\mathbf{W}=[\mathbf{w}(1),\mathbf{w}(2),\cdots,\mathbf{w}(N)].$ The problem to be addressed here is the estimation of the number of sources $K$, the channel matrix $\mathbf{H}$, and the symbol matrix $\mathbf{X}$, given the array output $\mathbf{Y}$.

\subsubsection{Estimation of the number of sources K}
For an easy start, we ignore the white noise for the moment and have $\mathbf{Y}=\mathbf{HX}$. The rank of $\mathbf{H}$ is equal to $K$ if $K<M$. Likewise, $\mathbf{X}$ is full-row-rank when $N$ is much larger than $K$. Consequently, we have $rank(\mathbf{Y})=K$ and $K$ is equal to the number of nonzero singular values of $\mathbf{Y}$. With white noise added to the data, $K$ can be estimated from the number of singular values of $\mathbf{Y}$ that are significantly larger than zero.

\subsubsection{Estimation of \textbf{X} and \textbf{H}}
In this paper, we adopt the Finite Alphabet (FA) based blind detection algorithm to estimate $\mathbf{X}$ and $\mathbf{H}$, assuming $K$ is known. The maximum-likelihood estimator yields the following separable least-squares minimization problem \cite{Talwar:96}
\begin{equation}\label{eqn:42}
    \min_{\mathbf{H},\mathbf{X}\in\Omega}\|\mathbf{Y}-\mathbf{HX}\|_F^2
\end{equation}
where $\Omega$ is the finite alphabet to which the elements of $\mathbf{X}$ belong, and $\|\cdot\|_F^2$ is the Frobenius norm. The minimization of (\ref{eqn:42}) can be carried out in two steps. First, we minimize (\ref{eqn:42}) with respect to $\mathbf{H}$ and obtain
\begin{equation}\label{eqn:43}
    \hat{\mathbf{H}}=\mathbf{YX}^\texttt{+}=\mathbf{YX}^H(\mathbf{XX}^H)^{-1},
\end{equation}
where $(\cdot)^+$ is the pseudo-inverse of a matrix. Substituting $\hat{\mathbf{H}}$ back into (\ref{eqn:42}), we obtain a new criterion, which is a function of $\mathbf{X}$ only:
\begin{equation}\label{eqn:44}
    \min_{\mathbf{X}\in\Omega}\|\mathbf{YP}_{\mathbf{X}^H}^\bot\|_F^2,
\end{equation}
where $\mathbf{P}_{\mathbf{X}^H}^\bot=\mathbf{I}-\mathbf{X}^H(\mathbf{XX}^H)^{-1}\mathbf{X}$, and $\mathbf{I}$ is the identity matrix. The global minimum of (\ref{eqn:44}) can be obtained by enumerating over all possible choices of $\mathbf{X}$. Reduced-complexity iterative algorithms that solve (\ref{eqn:44}) iteratively such as ILSP and ILSE were introduced in \cite{Papadias:97}. Not being one of the foci of this paper, the details of ILSP and ILSE are not covered here. Interested readers are referred to \cite{Talwar:94} and the references therein.

Note that the scheme proposed in this section is only one way of implementing MPR in WLANs. It ensures that the orthogonal training sequences are transmitted in the preambles of data packets. This leads to highly reliable channel estimation that facilitates the user of MUD techniques. Moreover, the modification to the original protocol is mainly restrained within the AP. Minimum amendment is needed at mobile stations.

\section{Discussions}

\subsection{Random channel error}
In our analysis so far, we have assumed that packet error rate due to random fading effect is negligible when the number of simultaneous transmission is smaller than \emph{M} and is close to 1 otherwise. This assumption is quite accurate when data packets are well protected by error correction codes (e.g., convolutional codes in IEEE 802.11 protocol) and linear MUD is deployed at the receiver. The simplification allows us to focus on the effect of MPR on WLAN without the need to consider signal processing details such as coding and detection schemes.

In this section, we relax the assumption and investigate how random channel errors would affect our analysis. Fortunately, we can prove that super-linear throughput scaling still holds even when random channel error is taken into account, as detailed in the following. Denote by $P_M^{err}(k)$ the packet error rate due to wireless channel fading when \emph{k} packets are transmitted at the same time in a network with MPR capability $M$. Then, $P_M(k)=1-P_M^{err}(k)$ is the packet success rate, which is the probability that a packet \emph{survives} random channel fading \cite{Proakis}. Typically, $P_M(k)\geq P_M(k')$ for $k\leq k'$ and $P_M(k)\geq P_{M'}(k)$ for $M\geq M'$. Assuming linear detectors, we have $P_M(k)\approx 0$ if $k>M$ and $P_M(M)\approx P_{M'}(M')$ for $M\neq M'$ \cite{Winters:94}.

For simplicity, assume $T_{slot}=T_s=T_i=T_c=L/R$. Then, asymptotic throughput is given by
\begin{eqnarray}
S_\infty(M,\lambda)&=&R\sum_{k=1}^Mk\frac{\lambda^{k}}{k!}e^{-\lambda}P_M(k)\nonumber\\
&=&R\sum_{k=0}^{M-1}\frac{\lambda^{k+1}}{k!}e^{-\lambda}P_M(k+1)
\end{eqnarray}

At the optimal $\lambda^*(M)$, $\frac{\partial S_\infty(M,\lambda)}{\partial \lambda}=0$. Consequently,
\begin{eqnarray}\label{eqn56}
&&\sum_{k=0}^{M-1}\frac{(\lambda^*(M))^k}{k!}e^{-\lambda^*(M)}P_M(k+1)\nonumber\\
&=&\sum_{k=0}^{M-2}\frac{(\lambda^*(M))^{k+1}}{k!}e^{-\lambda^*(M)}(P_M(k+1)-P_M(k+2))\nonumber\\
&+&\frac{(\lambda^*(M))^M}{(M-1)!}e^{-\lambda^*(M)}P_M(M)\nonumber\\
&\leq&\frac{(\lambda^*(M))^M}{(M-1)!}e^{-\lambda^*(M)}P_M(M)
\end{eqnarray}
We are now ready to prove super-linear throughput scaling  $\frac{S_\infty^*(M+1)}{M+1}\geq\frac{S_\infty^*(M)}{M}$ in the following.
\begin{eqnarray}
&&S_\infty^*(M+1)\geq S_\infty(M+1,\lambda^*(M))\nonumber\\
&=&R\sum_{k=0}^{M-1}\frac{\lambda^*(M)^{k+1}}{k!}e^{-\lambda^*(M)}P_{M+1}(k+1)\nonumber\\
&&+R\frac{\lambda^*(M)^{M+1}}{M!}e^{-\lambda^*(M)}P_{M+1}(M+1)\nonumber\\
&\geq&S_\infty(M,\lambda^*(M))+R\frac{\lambda^*(M)^{M+1}}{M!}e^{-\lambda^*(M)}P_{M}(M)\nonumber\\
&\geq& \frac{M+1}{M}S_\infty^*(M)
\end{eqnarray}
where the last inequality is due to (\ref{eqn56}).

\subsection{Near far effect}
One implicit assumption in our analysis is that each station transmits at the same data rate $R$. In practice, stations experience different channel attenuation to the AP due to their random locations. If stations transmit at the same power level, then the data rate sustainable on each link would differ. In this case, the airtime occupied by a busy period is dominated the lowest data rate involved. Hence, the effective throughput enjoyed by high-rate stations would degenerate to the level of the lowest rate. Such problem, known as ``performance anomaly", is not unique to MPR. It exists in all multi-rate IEEE 802.11 networks. Fortunately, performance anomaly only causes the data rate $R$ in our throughput expression to degrade to $R_{min}$, where $R_{min}$ is the lowest possible data rate. Therefore, it will not affect the scaling law of throughput in MPR networks.

\subsection{Comparison with multiuser SIMO systems}
In this paper, we have demonstrated the drastic increase in spectrum efficiency brought by MPR. To implement MPR, modification is needed in both MAC and PHY layers, as discussed in Section V. With the same hardware enhancement (e.g., having $M$ antennas at the AP), an alternative is to let each link transmit at a higher data rate, but keep the single-packet-reception restriction unchanged. This essentially becomes a traditional WLAN with SIMO (single-input-multiple-output) links.

The capacity of a SIMO link increases logarithmically with the number of antennas at the receiver \cite{Tse}. That is,
\begin{equation} \label{eqn59}
R_{SIMO}\approx R_{SISO}+\log(M)
\end{equation}
where $R_{SISO}$ is the data rate of a SISO (single-input-single-output) link. In contrast, the data rate $R$ of each link in MPR WLAN is set to $R_{SISO}$, for antenna diversity is used to separate multiple data streams instead of increasing the rate of one stream therein.

With (\ref{eqn59}), the throughput of WLAN with SIMO links is
\begin{equation}
S_N^{SIMO}=\frac{L\sum_{k=1}^Mk\Pr\{X=k\}}{P_{idle}^{SIMO}T_i^{SIMO}+P_{coll}^{SIMO}T_c^{SIMO}+P_{succ}^{SIMO}T_s^{SIMO}}
\end{equation}
where the expressions for $P_{idle}^{SIMO}$, $P_{coll}^{SIMO}$, and $P_{succ}^{SIMO}$ are the same as (\ref{eqn:9}), (\ref{eqn:10}), and (\ref{eqn:11}) with $M=1$, respectively. Likewise, $T_i^{SIMO}$, $T_c^{SIMO}$, and $T_s^{SIMO}$ are the same as (\ref{eqn:1}), (\ref{eqn:2}), or (\ref{eqn:3}) except that $R$ is replaced by $R_{SIMO}$. Specifically, throughput in the ALOHA case becomes
\begin{equation}
S_N^{SIMO}=(R+\log(M))Np_t(1-p_t)^{N-1}.
\end{equation}
and the optimal $p_t$ that maximizes the throughput is equal to $1/N$. In particular, the maximum achievable throughput when $N$ is large is
\begin{equation}
S_\infty^{SIMO*}(M)=(R+\log(M))e^{-1}.
\end{equation}

It is obvious that the normalized throughput $\frac{S_\infty^{SIMO*}(M)}{M}$ decreases with $M$ in SIMO networks. This, in contrast to the super-linear throughput scaling in MPR networks, suggests that multiple antennas at the AP should be used to resolve simultaneous transmissions instead of increasing per-link data rate in random access WLANs.

\section{Conclusion}
With the recent advances in PHY-layer MUD techniques, it is no longer a physical constraint for the WLAN channel to accommodate only one packet transmission at one time. To fully utilize the MPR capability of the PHY channel, it is essential to understand the fundamental impact of MPR on the MAC-layer. This paper has studied the characteristic behavior of random-access WLANs with MPR. Our analysis provides a theoretical foundation for the performance evaluation of WLANs with MPR, and it is useful for system design in terms of setting operating parameters of MAC protocols.

Our analytical framework is general and applies to various WLANs including non-carrier-sensing and carrier-sensing networks. In Theorems 1 and 3, we have proved that the throughput increases super-linearly with $M$ for both finite and infinite population cases. This is the case in non-carrier-sensing networks for all $M$, and in carrier-sensing networks for moderate to large $M$. Moreover, Theorem 2 shows that the throughput penalty due to distributed random access diminishes when $M$ approaches infinity. Such scalability provides strong incentives for further investigations on engineering and implementation details of MPR systems. Based on the analysis, we found that the commonly deployed BEB scheme is far from optimum in most systems except the carrier-sensing systems with RTS/CTS four-way handshake. In particular, the optimum backoff factor $r$ increases with $M$ for large $M$. We further note that the throughput degrades sharply when $r$ is smaller than the optimum value, while it is much less sensitive to $r$ when $r$ exceeds the optimum.

Having understood the fundamental behavior of MPR, we propose practical protocols to exploit the advantage of MPR in IEEE 802.11-like WLANs. By incorporating advanced PHY-layer blind detection and MUD techniques, the protocol can implement MPR in a fully distributed manner with marginal modification of MAC layer.


%
%
\appendices
\section{Super-Linear Throughput Scaling in WLANs with Finite Population}
\begin{theorem}
(Super-linearity with finite population) $S_N^*(M+1)\big/(M+1) \geq S_N^*(M)\big/M$ for all $M<N$

From (\ref{eqn:12}), we have
\begin{eqnarray}\label{eqn:A1}
  S_N(M,p_t)&=& R\sum_{k=1}^Mk\binom{N}{k}p_t^k(1-p_t)^{N-k} \nonumber\\
  &=& R\frac{Np_t}{1-p_t}\sum_{k=0}^{M-1}\binom{N}{k}p_t^k(1-p_t)^{N-k} \nonumber\\
  &&-R\frac{p_t}{1-p_t}\sum_{k=0}^{M-1}k\binom{N}{k}p_t^k(1-p_t)^{N-k}\nonumber \\
  &=&R\frac{Np_t}{1-p_t}\Pr\{X\leq M-1\}\nonumber\\
  &&-\frac{p_t}{1-p_t}S_N(M-1,p_t)
\end{eqnarray}
and
\begin{equation}\label{eqn:A2}
    S_N(M+1,p_t)=R\frac{Np_t}{1-p_t}\Pr\{X\leq M\}-\frac{p_t}{1-p_t}S_N(M,p_t).
\end{equation}

Meanwhile
\begin{eqnarray}\label{eqn:A3}
    S_N(M+1,p_t)&=&R\sum_{k=1}^{M+1}k\binom{N}{k}p_t^k(1-p_t)^{N-k}\nonumber\\
    &=&S_N(M,p_t)+R(M+1)\Pr\{X=M+1\}\nonumber\\
    &&
\end{eqnarray}
Substituting (\ref{eqn:A3}) to (\ref{eqn:A2}), we get
\begin{eqnarray}\label{eqn:A4}
    S_N(M,p_t)=RNp_t\Pr\{X\leq M\}\nonumber\\
    -R(1-p_t)(M+1)\Pr\{X=M+1\}\;\forall M<N, p_t
\end{eqnarray}

At the optimal $p_t^*(M)$, the derivative $\partial S_N(M,p_t)/\partial p_t=0$. Thus,
\begin{eqnarray}\label{eqn:A5}
    \frac{\partial S_N(M,p_t)}{\partial p_t}\bigg|_{p_t=p_t^*(M)}=RN\Pr\{X\leq M\}\big|_{p_t=p_t^*(M)}\nonumber\\
    +R(M+1)\big(1-\frac{M+1}{p_t}\big)\Pr\{X=M+1\}\big|_{p_t=p_t^*(M)}=0,
\end{eqnarray}
Combining (\ref{eqn:A4}) and (\ref{eqn:A5}),
\begin{eqnarray}\label{eqn:A6}
&&    S_N(M,p_t^*(M))=p_t^*(M)\frac{\partial S_N(M,p_t)}{\partial p_t}\bigg|_{p_t=p_t^*(M)}\nonumber\\
    &&-R(M+1)\big(p_t^*(M)-(M+1)\big)\Pr\{X=M+1\}\big|_{p_t=p_t^*(M)}\nonumber\\
    &&-R(M+1)\big(1-p_t^*(M)\big)\Pr\{X=M+1\}\nonumber\\
    &=&RM(M+1)\Pr\{X=M+1\}\big|_{p_t=p_t^*(M)}
\end{eqnarray}
It is obvious that
\begin{eqnarray}\label{eqn:A7}
    &&S_N(M+1, p_t^*(M+1))\geq S_N(M+1,p_t^*(M))\nonumber\\
    &&=S_N(M,p_t^*(M))+R(M+1)\Pr\{X=M+1\}\big|_{p_t^*(M)}\nonumber\\
    &&
\end{eqnarray}
Substituting (\ref{eqn:A6}) to (\ref{eqn:A7}), we have
\begin{eqnarray}\label{eqn:A8}
  &&S_N(M+1,p_t^*(M+1))\nonumber\\
   &\geq& S_N(M,p_t^*(M))+\frac{S_N(M,p_t^*(M))}{M} \nonumber \\
  &=&S_N(M,p_t^*(M))\frac{M+1}{M}.
\end{eqnarray}
Hence, $S_N^*(M+1)\big/(M+1) \geq S_N^*(M)\big/M$ for all $M<N$.
\begin{flushright}
$\Box$
\end{flushright}

\end{theorem}




%

%
\begin{biography}[{\includegraphics[width=1in,height=1.25in,clip,keepaspectratio]{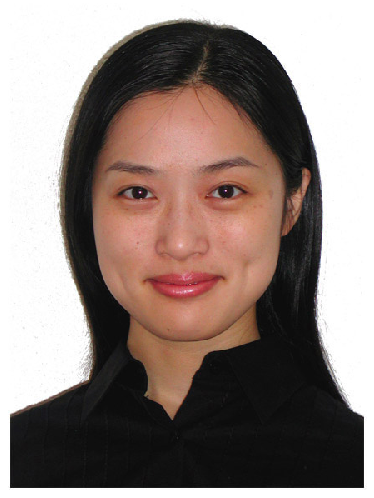}}]{Ying Jun (Angela) Zhang}
(S'00, M'05) received her BEng degree with Honors in Electronic Engineering  from Fudan University, Shanghai, China, and the PhD degree in Electrical and Electronic Engineering from the Hong Kong University of Science and Technology, Hong Kong. Since Jan. 2005, she has been with the Department of Information Engineering in The Chinese University of Hong Kong, where she is currently an assistant professor.

Dr. Zhang is on the Editorial Boards of IEEE Transactions of Wireless Communications and Willey Security and Communications Networks Journal. She has served as a TPC Co-Chair of Communication Theory Symposium of IEEE ICC 2009, Track Chair of ICCCN 2007, and Publicity Chair of IEEE MASS 2007. She has been serving as a Technical Program Committee Member for leading conferences including IEEE ICC, IEEE GLOBECOM, IEEE WCNC, IEEE ICCCAS, IWCMC, IEEE CCNC, IEEE ITW, IEEE MASS, MSN, ChinaCom, etc. Dr. Zhang is an IEEE Technical Activity Board GOLD Representative, 2008 IEEE GOLD Technical Conference Program Leader, IEEE Communication Society GOLD Coordinator, and a Member of IEEE Communication Society Member Relations Council.

Her research interests include wireless communications and mobile networks, adaptive resource allocation, optimization in wireless networks, wireless LAN/MAN, broadband OFDM and multicarrier techniques, MIMO signal processing.
As the only winner from Engineering Science, Dr. Zhang has won the Hong Kong Young Scientist Award 2006, conferred by the Hong Kong Institution of Science.
\end{biography}

\begin{biography}[{\includegraphics[width=1in,height=1.25in,clip,keepaspectratio]{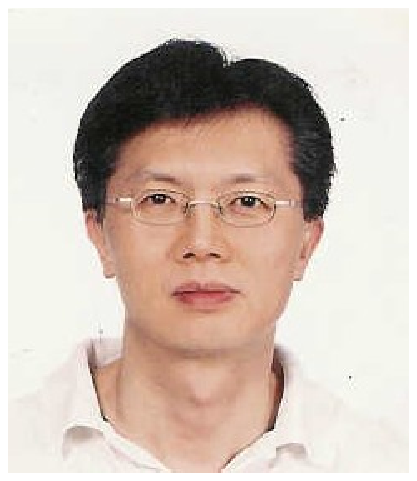}}]{Soung Chang Liew} (S'84-M'87-SM'92)
received his  S.B., S.M., E.E., and Ph.D. degrees from the Massachusetts Institute of Technology. From 1984 to 1988, he was at the MIT Laboratory for Information and Decision Systems, where he investigated Fiber-Optic Communications Networks. From March 1988 to July 1993, Soung was at Bellcore (now Telcordia), New Jersey, where he engaged in Broadband Network Research. Soung is currently Professor and Chairman of the Department of Information Engineering,  the Chinese University of Hong Kong. He is Adjunct Professor at Southeast University, China.

Soung's current research interests include wireless networks, Internet protocols, multimedia communications, and packet switch design. Soung and his student won the best paper awards in the 1st IEEE International Conference on Mobile Ad-hoc and Sensor Systems (IEEE MASS 2004) the 4th IEEE International Workshop on Wireless Local Network (IEEE WLN 2004). Separately, TCP Veno, a version of TCP to improve its performance over wireless networks proposed by Soung and his student, has been incorporated into a recent release of Linux OS. In addition, Soung initiated and built the first inter-university ATM network testbed in Hong Kong in 1993.

Besides academic activities, Soung is also active in the industry. He co-founded two technology start-ups in Internet Software and has been serving as consultant to many companies and industrial organizations. He is currently consultant for the Hong Kong Applied Science and Technology Research Institute (ASTRI), providing technical advice as well as helping to formulate R\&D directions and strategies in the areas of Wireless Internetworking, Applications, and Services.

Soung is the holder of four U.S. patents and Fellow of IEE and HKIE. He is listed in Marquis Who's Who in Science and Engineering. He is the recipient of the first Vice-Chancellor Exemplary Teaching Award at the Chinese University of Hong Kong. Publications of Soung can be found in www.ie.cuhk.edu.hk/soung.
\end{biography}

\begin{biography}[{\includegraphics[width=1in,height=1.25in,clip,keepaspectratio]{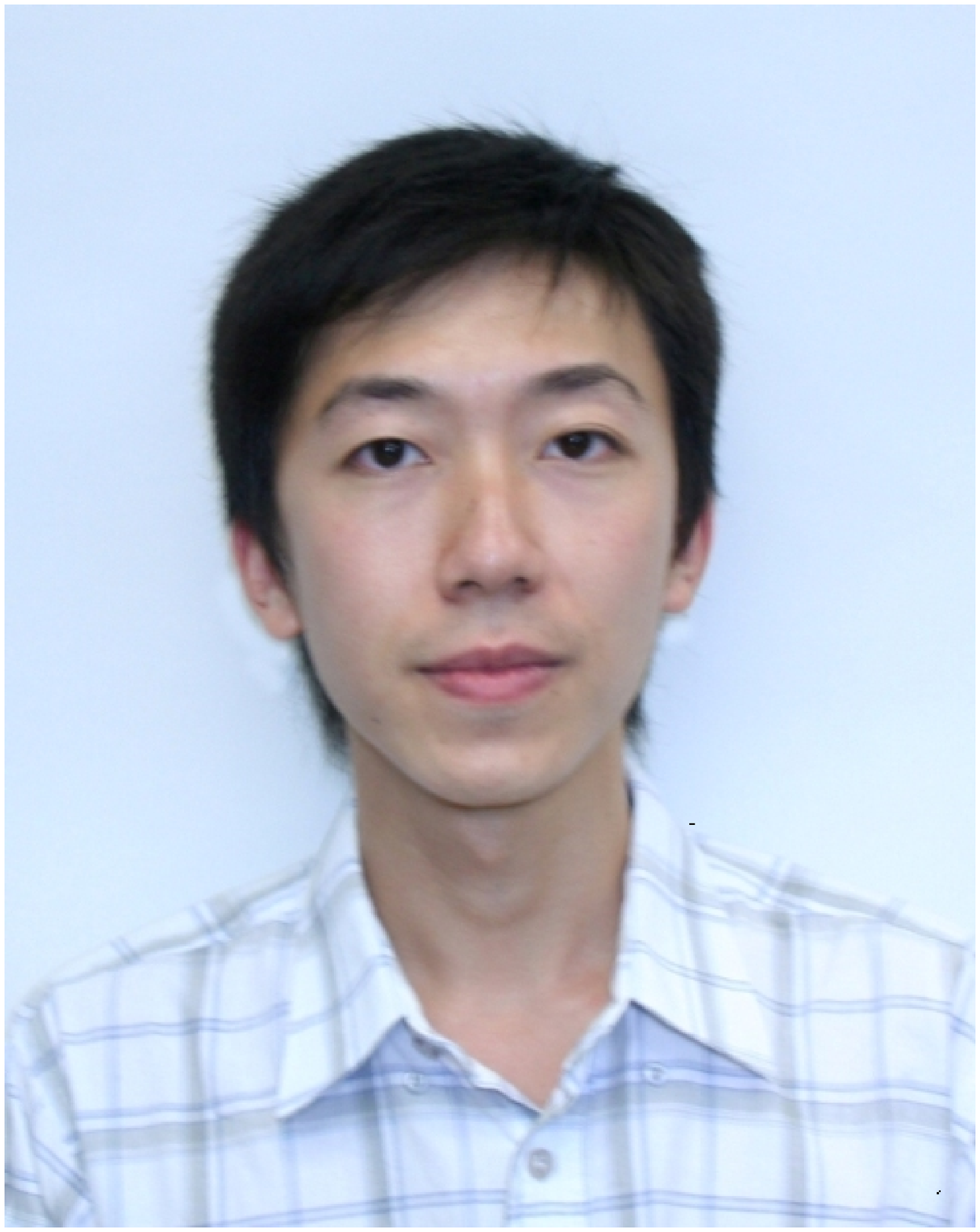}}]{Pengxuan Zheng} (S'06) received his B.Eng. degree in Computer Science and Engineering from Zhejiang University, Hangzhou, China, in 2004, and M.Phil. degree in Information Engineering from the Chinese University of Hong Kong, Hong Kong, China, in 2007. He is currently working towards the Ph.D. degree at Purdue University. His research interests include mobile ad hoc networks and wireless sensor networks.
\end{biography}


%
%
%
%
%
%
%

\end{document}